\documentclass[10pt,aps,prd,nofootinbib,superscriptaddress,twocolumn,preprintnumbers,balancelastpage]{revtex4}

\usepackage[a-2b,mathxmp]{pdfx}

\usepackage{graphicx,epsfig,psfrag,amssymb} %,hyperref}
\usepackage{multirow}
\usepackage{xcolor,graphicx,epsfig,psfrag,amsmath,empheq}
\usepackage{bm}
\usepackage{mathrsfs,amsfonts,color}
\usepackage{slashed}
\usepackage{hepunits}
\usepackage{color}
\usepackage{enumitem}

\usepackage[utf8]{inputenc}
\newcommand{\beq}{\begin{equation}}
\newcommand{\eeq}{\end{equation}}
\newcommand{\beqa}{\begin{eqnarray}}
\newcommand{\eeqa}{\end{eqnarray}}

\def\lsim{\mathrel{\rlap{\lower4pt\hbox{\hskip1pt$\sim$}}
    \raise1pt\hbox{$<$}}}         %less than or approx. symbol
\def\gsim{\mathrel{\rlap{\lower4pt\hbox{\hskip1pt$\sim$}}
    \raise1pt\hbox{$>$}}}         %greater than or approx. symbol

\renewcommand{\keV}{{\rm keV}}

\renewcommand{\GeV}{{\rm GeV}}
\renewcommand{\TeV}{{\rm TeV}}

\newcommand{\cL}{\mathcal{L}}

\newcommand{\cM}{\mathcal{M}}
\newcommand{\cR}{\mathcal{R}}

\newcommand{\abs}[1]{\left| #1 \right|}

\newcommand{\XeT}{{XENON1T}\xspace}

\begin{document}
%\widetext

%\preprint{}

%%%%%%%%%%%%%%%%%%%%%%%%%%%%%%%%%%%%%%%%
\title{Probing the relaxed relaxion and Higgs-portal with S1 \& S2} 
%%%%%%%%%%%%%%%%%%%%%%%%%%%%%%%%%%%%%%%%

\author{Ranny Budnik}
\email{ran.budnik@weizmann.ac.il}
\affiliation{Department of Particle Physics and Astrophysics, Weizmann Institute of Science, Rehovot 7610001, Israel}
\affiliation{Simons Center for Geometry and Physics and C. N. Yang Institute for Theoretical Physics, SUNY, Stony Brook, NY, USA}

\author{Hyungjin Kim}
\email{hyungjin.kim@weizmann.ac.il}
\affiliation{Department of Particle Physics and Astrophysics, Weizmann Institute of Science, Rehovot 7610001, Israel}

\author{Oleksii Matsedonskyi}
\email{oleksii.matsedonskyi@weizmann.ac.il}
\affiliation{Department of Particle Physics and Astrophysics, Weizmann Institute of Science, Rehovot 7610001, Israel}

\author{Gilad Perez}
\email{gilad.perez@weizmann.ac.il}
\affiliation{Department of Particle Physics and Astrophysics, Weizmann Institute of Science, Rehovot 7610001, Israel}

\author{Yotam Soreq}
\email{soreqy@physics.technion.ac.il}
\affiliation{Physics Department, Technion---Israel Institute of Technology, Haifa 3200003, Israel}

%%%%%%%%%%%%%%%%%%%%%%%%%%%%%%%%%%%%%%%%
\begin{abstract}
\noindent
We study the recent \XeT excess in context of solar scalar, specifically in the framework of Higgs-portal and the relaxion model.
We show that $m_\phi = 1.9\,\keV$ and $g_{\phi e}=2.4\times 10^{-14}$ can explain the observed excess in science run 1 (SR1) analysis in the 1-7 keV range. 
When translated into the scalar-Higgs mixing angle, the corresponding mixing angle $\sin\theta = 10^{-8}$ is intriguingly close to the maximum value of mixing angle for the technical naturalness of the scalar mass. 
Unlike the solar axion model, the excess favors a massive scalar field because of its softer spectrum. 
In the minimal scenarios we consider, the best fit parameters are in tension with stellar cooling bounds. 
We discuss a possibility that a large density of red giant stars may trigger a phase transition, resulting in a local scalar mass increase suppressing the stellar cooling. 
For the particular case of minimal relaxion scenarios, we find that such type of chameleon effects is automatically present but they can not ease the cooling bounds. 
They are however capable of triggering a catastrophic phase transition in the entire universe. 
Following this observation 
we derive a new set of bounds on the relaxed-relaxion parameter space.
\end{abstract}
%%%%%%%%%%%%%%%%%%%%%%%%%%%%%%%%%%%%%%%%

%\pacs{}

\maketitle

%%%%%%%%%%%%%%%%%%%%%%%%%%%%%%%%%%%%%%%%
\section{Introduction}
\label{sec:intro}
%%%%%%%%%%%%%%%%%%%%%%%%%%%%%%%%%%%%%%%%

Recently, the \XeT experiment reported an excess of electronic recoil events in the SR1 signal~\cite{Aprile:2020tmw}.
Within the energy range of $1-7\,\keV$, the expected number of background only events is $232\pm15$, while the observed number of events is 285 with an apparent peak near $2-3\,\keV$, in contrast to the expected flat background. 
The discrepancy corresponds to a $3.5\,\sigma$ rejection of the background hypothesis in favor of an additional peaked spectrum resembling a solar axion source~\cite{Aprile:2020tmw}. 
An unaccounted-for background of tritium decay would lower the significance of the excess to about $2.2\,\sigma$.  
While it is possible that the excess is due to a statistical fluctuation or yet another unaccounted background, we focus on the case that it is due to the existence of a new degree of freedom with a mass smaller than a few keV.

The interpretation for the excess as a solar axion with $m_a\lesssim 0.1\,\keV$ leads to an electronic coupling of $g_{ae}\sim 3\times10^{-12}$, where the corresponding upper bound is $g_{ae}<3.7\times 10^{-12}$~\cite{Aprile:2020tmw}.
This is consistent with the LUX solar axion search, which implies an upper bound of $g_{ae}<3.5\times10^{-12}$~\cite{Akerib:2017uem}, but in tension with astrophysical bounds from stellar cooling.
Ref.~\cite{Viaux:2013lha} reported an upper bound of $g_{ae}\lesssim3\times 10^{-13}$ from red giant~(RG) stars cooling.
Yet, there are hints for a signal in anomalous energy loss in white dwarfs, RG stars and neutron stars which point to a preferred coupling of $g_{ae}=(1.6\pm0.3)\times10^{-13}$~\cite{Giannotti:2017hny} (see also~\cite{Hansen:2015lqa}). 
However, as pointed by Ref.~\cite{Budnik:2019olh}, a light scalar with a mass at or below the keV scale can be produced in the Sun and be probed by dark matter~(DM) direct detection experiments through electron ionisation at the keV scale. 
In this work we mainly focus on this possibility and confront it with the \XeT excess. Other possible implications of the recent XENON1T data were discussed in~\cite{Takahashi:2020bpq,Kannike:2020agf,Alonso-Alvarez:2020cdv,Amaral:2020tga,Fornal:2020npv,Boehm:2020ltd,Bally:2020yid,Harigaya:2020ckz,Su:2020zny,Du:2020ybt,DiLuzio:2020jjp,Dey:2020sai,Chen:2020gcl,Bell:2020bes,Buch:2020mrg,Choi:2020udy,AristizabalSierra:2020edu,Paz:2020pbc,Lee:2020wmh,1802687,Cao:2020bwd,Primulando:2020rdk,Khan:2020vaf,Nakayama:2020ikz,1802726,1802727,1802729}.

Producing a light scalar (or a pseudo-scalar with CP-odd couplings) is generically a non-trivial task from the model-building point of view. We will concentrate on two cases: a generic scalar Higgs-portal scenario, which can be seen as an effective description of various more complicated constructions, and a more predictive relaxion model~\cite{Graham:2015cka} motivated by the Higgs mass naturalness problem. While the relaxion is considered to be a pseudo-scalar, its vacuum generically breaks CP~\cite{Flacke:2016szy,Choi:2016luu} leading to a scalar-like phenomenology.

Below, we analyse the recent science run 1 \XeT result~\cite{Aprile:2020tmw} in the context of a new scalar field with mass at the keV scale or below and show that such a new particle is compatible with the excess. 
In addition, we explore its implications for the S2-only analysis~\cite{Aprile:2019xxb}, and show that such scalar with keV mass has a clear signature in terms of a bump on flat background. 
However, the bounds from stellar cooling are stronger~\cite{Hardy:2016kme} and exclude the preferred region of the parameter space. 
We also consider the case where the tritium background is taken into account and show that the preferred parameter space is consistent with a smaller coupling and the tension with stellar cooling bound is weakened.
Finally, we map the relevant parameter space to the generic Higgs portal and the relaxion model~\cite{Graham:2015cka}. 
We also discuss a possibility that phase transition takes place inside RGs, locally increasing the scalar mass, and hence, alleviating the tension with stellar cooling bounds. 
We find that such phase transition occurs automatically in minimal (non-QCD) relaxion models for a finite region of parameter space, but unfortunately, we find that it does not ease the tension since the new phase within the dense stars is expected to expand rather than be localised. 
Nevertheless, we derive a new constraint on the relaxion parameter space, which is required to avoid a new phase, with unrealistic Higgs mass, to fill the whole Universe.

%%%%%%%%%%%%%%%%%%%%%%%%%%%%%%%%%%%%%%%%
\section{The solar relaxion/scalar signal}
\label{sec:signal}
%%%%%%%%%%%%%%%%%%%%%%%%%%%%%%%%%%%%%%%%

We estimate the solar scalar signal by following Ref.~\cite{Budnik:2019olh}. For the axion case, see~\cite{Redondo:2013wwa}.
The relevant $\phi$-electron interaction Lagrangian is given by
\begin{align}
	\cL 
	\supset
	-g_{\phi e} \phi\bar{e}e \, .
\end{align}
Below, we focus on the $m_\phi \lesssim 3\,\keV$ mass range and consider finite mass effects.

Within the Sun, light scalars can be produced by various production mechanisms:
bounded electrons~(bb), recombination of free electrons~(bf), Bremsstrahlung emission due to scatterings of electrons on ions~(ff), Bremsstrahlung emission due to scatterings of two electrons~(ee) and Compton like processes~(C).
At the relevant energy scale, the dominant production rate is the electron-ion Bremsstrahlung.
The total differential scalar flux is estimated as
\begin{align}
	\label{eq:PhiScalar}
	\frac{d\Phi}{d\omega}
	\approx
	\frac{\omega k}{8\pi^3 R^2}\int_{\odot} dV \ \Gamma^{\rm prod}(\omega)\, ,
\end{align}
where $\Gamma^{\rm prod}$ is sum over all production rates, $R=1\,{\rm AU}$ is the distance between the earth and the Sun, $\omega$ and $k$ are the scalar energy and momentum, respectively and $V$ is the Sun volume, where the Sun profile is taken from~\cite{Vinyoles:2016djt}. 

The ratio between the matrix elements of a $\gamma$ emission and a $\phi$ emission (or absorption) is given by~\cite{Avignone:1986vm}
\begin{align}
	\label{eq:M2ratio}
	\frac{\abs{\cM(e\to e\phi)}^2}{\abs{\cM(e\to e\gamma)}^2}
	\approx
	\frac{g^2_{\phi e}}{4\pi \alpha} \beta^2 \, ,
\end{align}
where $\beta=k/\omega$ is the scalar velocity. 
Since the ratio in Eq.~\eqref{eq:M2ratio} enters both in the production and in the detection (divided by $\beta$), the ratio between the number of scalar and pseudo scalar events rates can be written as 
\begin{align}
	\label{eq:Raphi}
	\frac{\cR_{\phi}(\omega)}{\cR_{a}(\omega)}
=	\frac{g^4_{\phi e}}{g^4_{a e} }  
	\frac{m^4_e}{ \omega^4}\left( 4 \, \frac{\beta^2}{3-\beta^{2/3}}\right)^2 \, ,
\end{align}
where $\cR_{\phi,a}(\omega) = \frac{d\Phi}{d\omega} \sigma_{\phi,a}(\omega)$, and $\sigma_{\phi,a}(\omega)$ is scalar and pseudo scalar absorption cross-section for liquid Xenon~\cite{Avignone:1986vm, Budnik:2019olh, Alessandria:2012mt, Pospelov:2008jk}. From Eq.~\eqref{eq:Raphi} we learn that the solar scalar signal is softer than the solar axion-like case, and thus, it will be peaked at lower energies. 

\begin{figure}[!t]
\vspace{-0.1cm}
\includegraphics[width=0.4\textwidth]{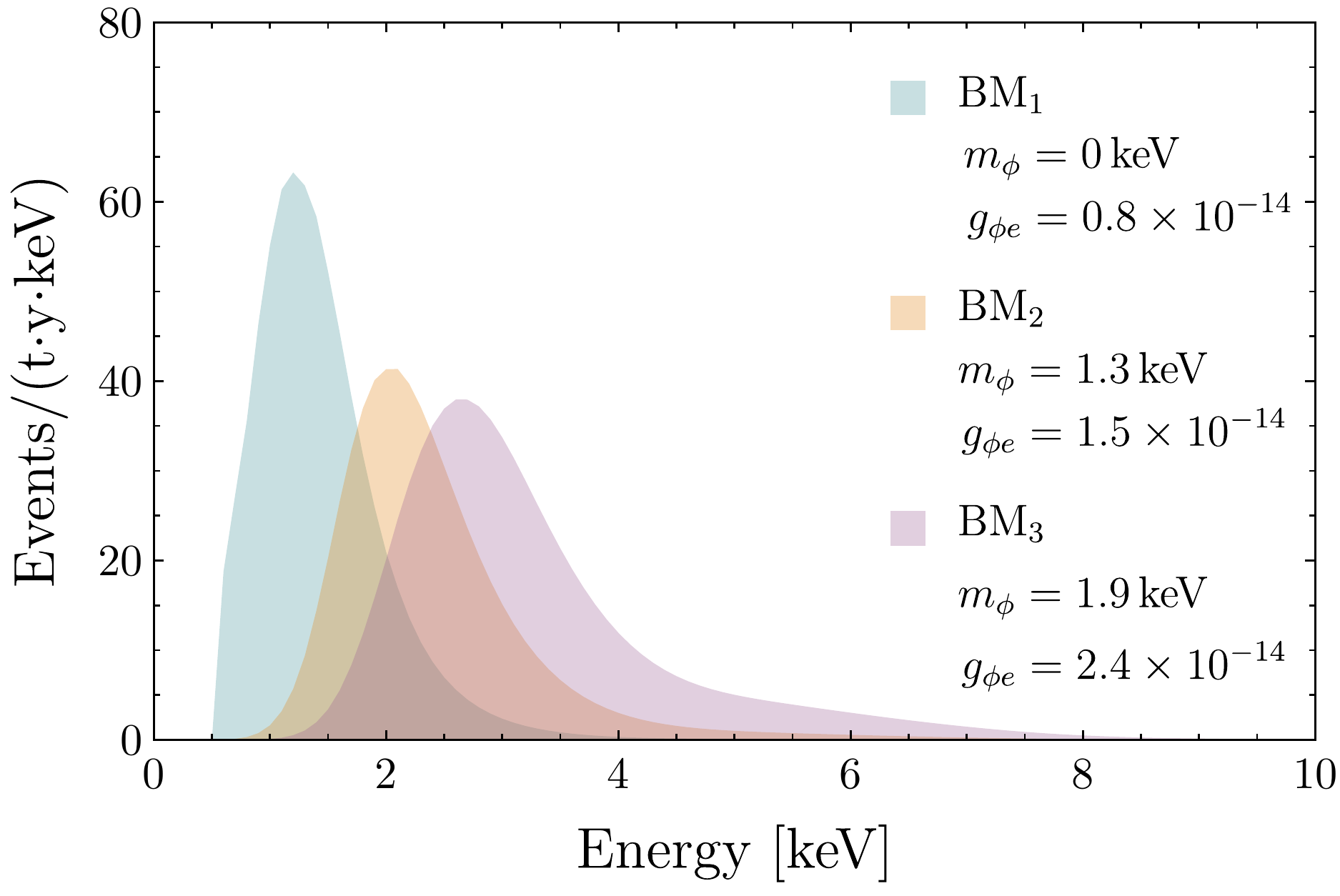}
 \vspace{-0.2cm}
\caption{
The solar scalar event rates are shown for three benchmark points as indicated on the plot. 
The shown event rates include the detector efficiency and resolution. See the main text for details.
}
 \vspace{-0.2cm}
\label{fig:RphiSpec}
\end{figure}

Next, following~\cite{Budnik:2019olh} we evaluate the $\cR_\phi(\omega) \epsilon_{\rm Xe}$, where the \XeT detector efficiency, $\epsilon_{\rm Xe}$, is taken from~\cite{Aprile:2020tmw}.  
The detector effects are taken into account by a Gaussian smearing of the signal, where the relevant parameters are adopted from~\cite{Aprile:2020yad}. 
The predicted $\phi$ event rates (after smearing) for three benchmark points, BM$_{1,2,3}$, with $m_{\phi}=(0,\,1.3,\,1.9)\,\keV$ and $g_{\phi e}=(0.8,\,1.5,\,2.4)\times 10^{-14}$, respectively, are plotted in Fig.~\ref{fig:RphiSpec}. 
We validated the smearing procedure by smearing the massless axion signal spectrum from Ref.~\cite{Redondo:2013wwa} and comparing it to Fig.~1 of~\cite{Aprile:2020tmw}, and found a matching up to few percent level. 
%%%%%%%%%%%%%%%%%%%%%%%%%%%%%%%%%%%%%%%% 
%%%%%%%%%%%%%%%%%%%%%%%%%%%%%%%%%%%%%%%%

In addition to the above signal, manifested in both a scintillation signal~(S1) and an ionisation signal~(S2), we consider the scalar signal in the XENON1T's  S2-only analysis~\cite{Aprile:2019xxb}, where the energy threshold is lower, $\sim 200\,\eV$.  
Since in the scalar case the signal is softer, it is expected to have a better sensitivity in the S2-only analysis. 
Scalars with masses around the solar interior plasma frequencies, $1\,\eV \lesssim \omega_{\rm p}\lesssim 300\,\eV$, have enhanced production rate due to mixing with the photon longitudinal mode in the Sun plasma~\cite{Hardy:2016kme}.
As pointed out in~\cite{Budnik:2019olh}, the resulting sensitivity for $m_\phi\lesssim300\,\eV$ by using the \XeT S2-only dataset~\cite{Aprile:2019xxb} is improved by an order of magnitude. 
This resonant production is only efficient for scalar masses below the local plasma frequency, which affects the shape of the expected spectrum with respect to the scalar mass. We finally note that in-medium mixing effect at the detector is negligible for the solar scalar, while it could be important for direct detection experiments for light scalar dark matter~\cite{Gelmini:2020xir}.

%%%%%%%%%%%%%%%%%%%%%%%%%%%%%%%%%%%%%%%%
\section{Recasting ot the \XeT excess as a relaxion/scalar}
\label{sec:scalar}
%%%%%%%%%%%%%%%%%%%%%%%%%%%%%%%%%%%%%%%%

We fit the scalar signal for the SR1 dataset of Ref.~\cite{Aprile:2020tmw} with and without the tritium background as follows.
In the first case, we construct a likelihood function of $m_\phi$ and $g_{\phi e}$ for the scalar signal and background. 
We take the background model as fixed, directly from Fig.~4 of~\cite{Aprile:2020tmw}.
This is justified as the background (without the tritium) is essentially fixed by the high energy spectrum and the injection of the signal at low energy has negligible effect on it. This is evident from the $\sim2\%$ change at the high end energy tail in the best fit while considering the solar axion, tritium and $\nu$ magnetic moment in  Ref.~\cite{Aprile:2020tmw}.
To assess the sensitivity of the result, we also add the tritium background component, where we profile over its magnitude.  

%%%%%%%%%%%%%%%%%%%%%%%%%%%%%%%%%%%%%%%% 
\begin{figure*}[t]
\vspace{-0.1cm}
\includegraphics[width=0.45\textwidth]{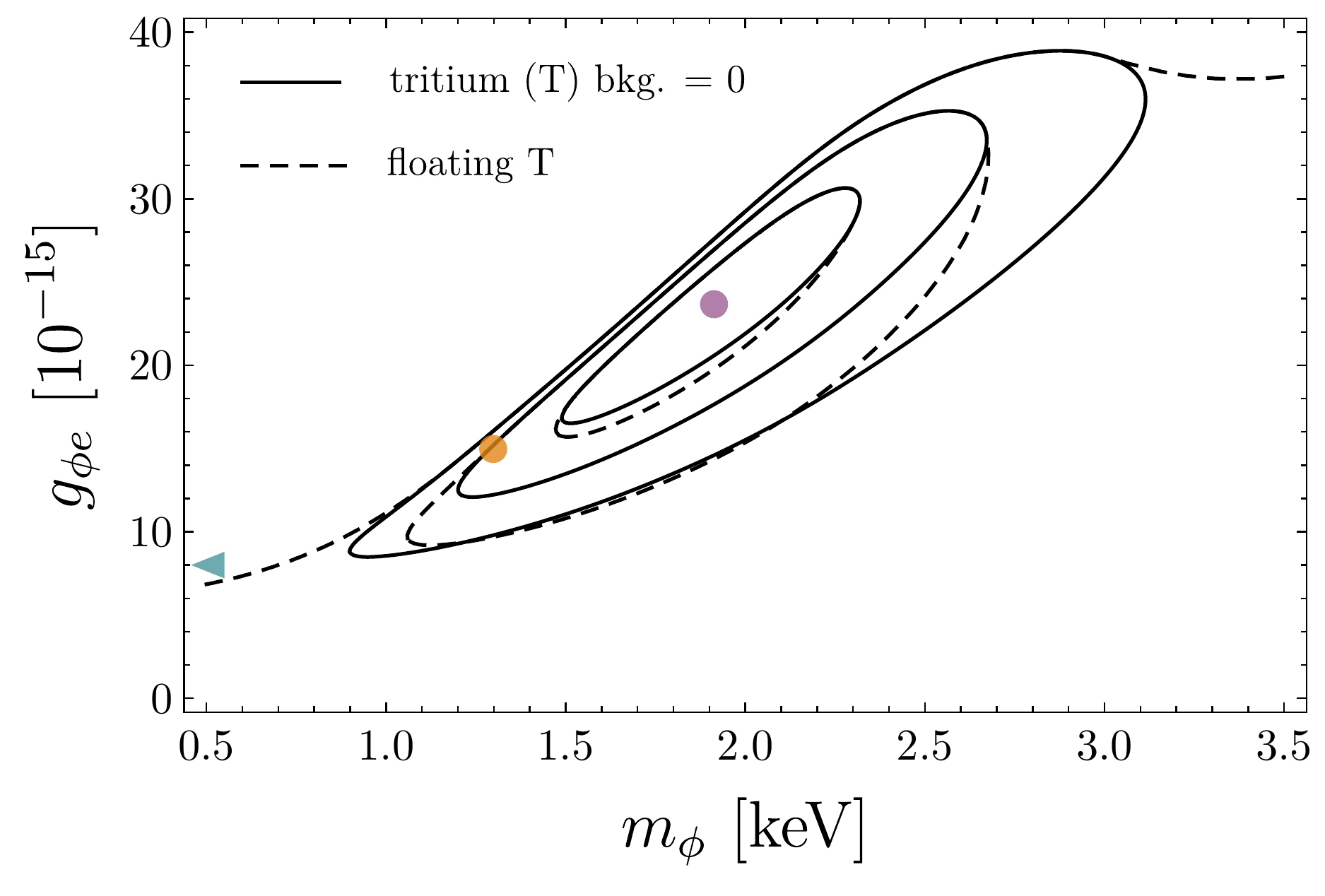}~~~~~~~~
\includegraphics[width=0.45\textwidth]{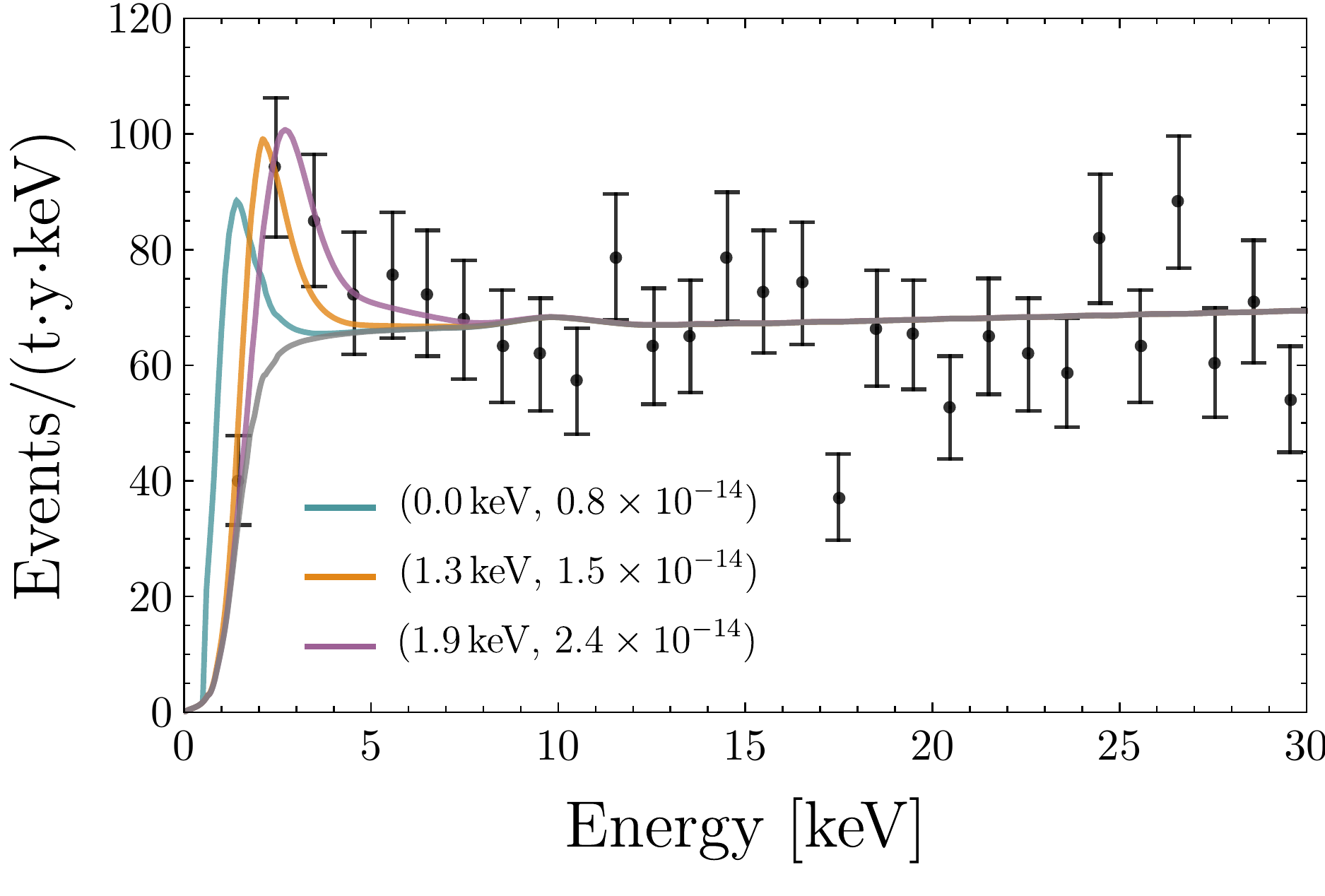}
 \vspace{-0.2cm}
\caption{
Left: 
The 68\,\%, 95\,\% and 99\,\% confidence intervals are shown, solid\,(dashed) contours are with\,(without) the tritium background.
The three benchmark points with $m_{\phi}=(0,\,1.3,\,1.9)\,\keV$ and $g_{\phi e}=(0.8,\,1.5,\,2.4)\times 10^{-14}$, are marked in cyan, orange and purple, respectively. The purple (BM$_3$) is the best fit point. 
Right:  
The signal+background is shown for the three benchmark points. 
The black points and gray line are data and background (without tritium) from~\cite{Aprile:2020tmw}, respectively. 
}
 \vspace{-0.2cm}
\label{fig:fitmphige}
\end{figure*} 
%%%%%%%%%%%%%%%%%%%%%%%%%%%%%%%%%%%%%%%%
By minimize the likelihood, the best fit point (with and without the tritium background) is found to be $m_\phi=1.9\,\keV$ and $g_{\phi e}=2.4\times 10^{-14}$, where the left panel of Fig.~\ref{fig:fitmphige} shows the 68\,\%, 95\,\% and 99\,\% confidence intervals in the $m_\phi-g_{\phi e}$ plane with and without the tritium background. 
To find the contours we apply the asymptotic formula from~\cite{Cowan:2010js} for two free parameters.
We find that the preferred region is for a finite $m_\phi\sim2\,\keV$.  
This is in contrast to the pseudo scalar case, where an effectively massless solution is favored.  
The reason  is that a massless or very light scalar spectrum has a significant soft component relative to the pseudo scalar case, as emphasised in  Eq.~\eqref{eq:Raphi}.
The right panel of Fig.~\ref{fig:fitmphige} demonstrates this point, showing a comparison between the signal and background with respect to the \XeT data for the three benchmark models, BM$_{1,2,3}$.
We note that the preferred region in the parameter space is in tension with the upper bound found from limits on RG cooling including plasmon-scalar mixing effect, $g_{\phi e}< 7 \times 10^{-16}$~\cite{Hardy:2016kme}. 
As a cross check, we have preformed the same likelihood analysis for the pseudoscalar case and found good agreement with the result of~\cite{Aprile:2020tmw}. 

\begin{figure}[t]
\vspace{-0.1cm}
\includegraphics[width=0.45\textwidth]{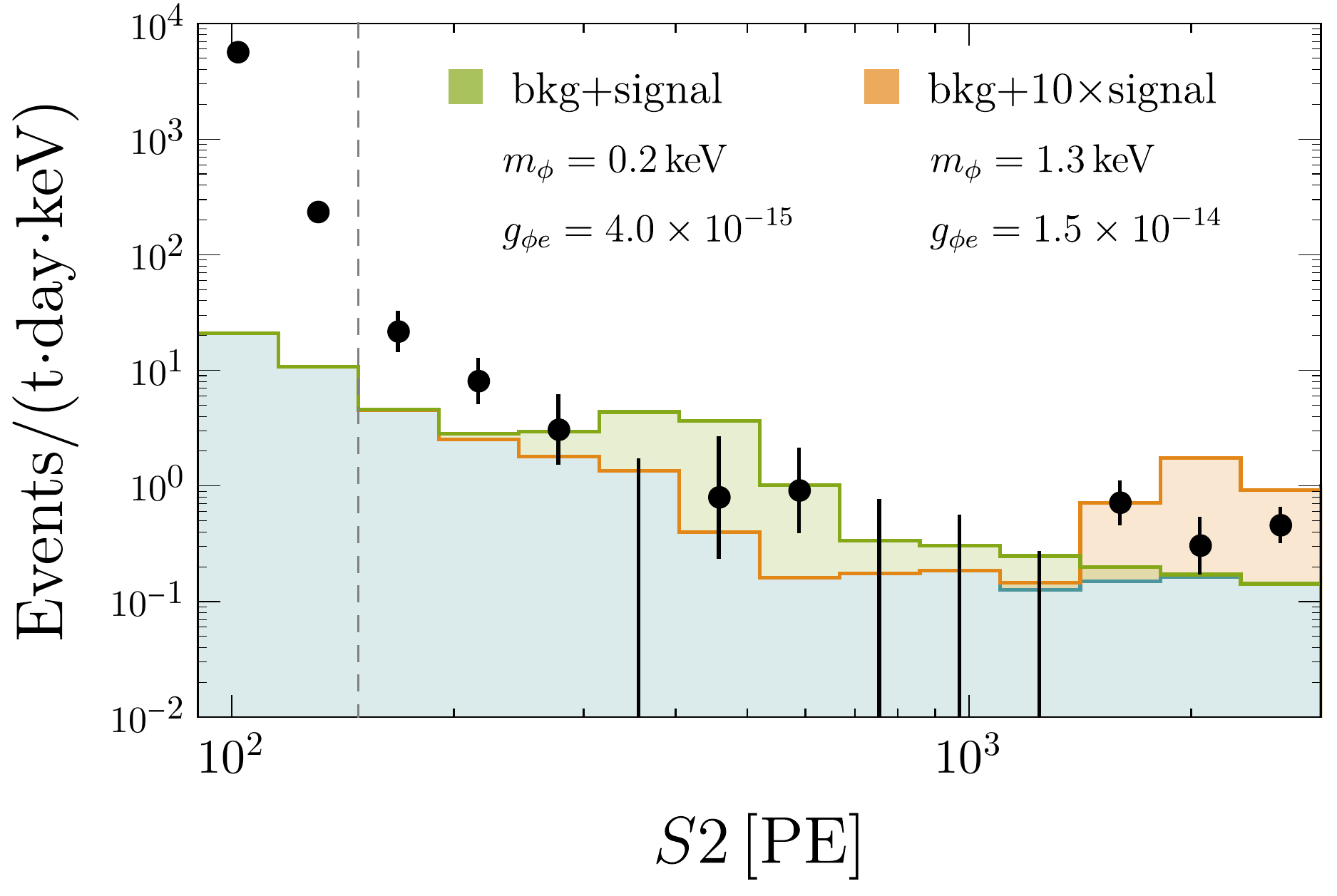}
 \vspace{-0.2cm}
\caption{
The signal and background for the S2-only analysis is shown. 
The BM$_2$ (orange), $m_\phi = 1.3\,\keV$ and $g_{\phi e} = 1.5\times 10^{-14}$, is chosen for this plot. The signal is enhanced by $10$ for the illustrative purpose. In addition, we have also plotted the signal that would arise from $m_\phi = 200\,\eV$ and $g_{\phi e} = 4 \times 10^{-15}$ (green). Here, the coupling constant is chosen to show the spectrum of events although it is only marginally consistent with the current S2-only analysis~\cite{Budnik:2019olh}.
Also, for this choice of mass, the event spectrum for SR1 excess is more or less similar to BM$_1$ for the same coupling constant. 
See the main text, especially Sec.~\ref{sec:relaxed}, for the potentially interesting phenomenological consequences related to this choice of parameters. 
}
 \vspace{-0.2cm}
\label{fig:S2Signal}
\end{figure}

In addition to the S1 and S2 signal, we now consider the possibility of a scalar  signal in the S2-only analysis of \XeT~\cite{Aprile:2019xxb}. This analysis only uses a partial background model, making possible setting upper bounds on signals and testing the consistency of a given signal.
In Fig.~\ref{fig:S2Signal}, we plot the S2-only expected signal for BM$_2$, $m_\phi  = 1.3\,\keV$ and $g_{\phi e} = 1.5\times 10^{-14}$, and compare it to the expected background and the data from~\cite{Aprile:2019xxb}.  
For the purpose of demonstration, we have multiplied  the signal by 10. 
We have also verified that the best fit parameters for SR1 dataset of Ref.~\cite{Aprile:2019xxb} is consistent with S2-only analysis. In addition to BM$_2$, we have also plotted the signals of $m_\phi = 200\,\eV$ and $g_{\phi e} = 4\times 10^{-15}$. 
For these parameters, the spectral shape of events for SR1 excess is close to the BM$_1$ in Fig.~\ref{fig:fitmphige}, while the events at the peak are suppressed by less than ten percent for the same coupling constant. 
This choice of parameters, especially in the context of relaxed relaxion, may lead to interesting phenomenological consequences inside stellar objects due to finite density corrections to the potential. This will be briefly discussed in Sec.~\ref{sec:relaxed}.
%%%%%%%%%%%%%%%%%%%%%%%%%%%%%%%%%%%%%%%% 
%%%%%%%%%%%%%%%%%%%%%%%%%%%%%%%%%%%%%%%%

%%%%%%%%%%%%%%%%%%%%%%%%%%%%%%%%%%%%%%%%
\section{Naturalness miracle}
%%%%%%%%%%%%%%%%%%%%%%%%%%%%%%%%%%%%%%%%

We will now confront the observed excess of events with  theoretical models. 
Let us start with the case of a generic scalar Higgs portal model, containing one new scalar degree of freedom. Its coupling to the electrons comes from the mixing with the Higgs and is given by
\begin{align}
	g_{\phi e} 
= 	\frac{\lambda_e}{\sqrt{2}} \sin \theta \,,
\end{align}
where $\lambda_e$ is the electron Yuakawa, and  $\sin \theta$ is the mixing whose best-fit value turns out to be $\sin \theta \simeq 1.2 \times 10^{-8}$.
In the absence of any special cosmological dynamics, naturalness implies an upper bound on the $\phi-h$ mixing angle (the red line in Fig.~\ref{fig:mixmass})~\cite{Piazza:2010ye,Arvanitaki:2015iga,Graham:2015ifn}:
\begin{align} 
	\label{eq:natbound}
	\sin \theta 
< 	\frac{m_\phi}{m_h} \simeq 1.5 \times 10^{-8}\left ( \frac{m_\phi}{1.9\, \keV} \right)\,,
\end{align}
where in the last equality we used the best-fit value for the scalar mass and $m_h\approx125\,\GeV$ is the Higgs mass.
While the best fit mixing satisfies the naturalness bound, it appears to be strikingly close to the boundary of the natural region. 
Below we will argue that in fact, in Higgs portal models, having a mixing close to the naturalness bound is much more likely than the mixing taking any given value far below the naturalness bound. 
 
Let us consider a generic Higgs portal potential, featuring a mass mixing term $\mu |H|^2 \phi$,  where $\mu$ sets the mixing strength. 
Because of this mixing, the scalar $\phi$ inherits the Higgs couplings, suppressed by the mixing angle 
\begin{align}
	\sin \theta \simeq \mu/v\,,
\end{align}
where $v$ is a Higgs VEV.
We also find the following physical $\phi$ mass around $\phi=0$
\begin{align}
	\label{eq:phimass}
	m_\phi^2 
= 	m_{\phi0}^2 - m_h^2 \sin^2 \theta\,,
\end{align}
where $m_{\phi0}^2$ is the bare $\phi$ mass. 
The \XeT excess corresponds to $m_\phi^2 \simeq m_h^2 \sin^2 \theta$. Such a relation can be reproduced for $m_{\phi0}^2 \sim m_h^2 \sin^2 \theta$, corresponding to $m_{\phi 0}\sim \mu$, but also for any $m_{\phi0}^2 \ll m_h^2 \sin^2 \theta$, corresponding to $m_{\phi_0} \ll \mu$. In the latter case the $\phi=0$ point is a maximum of the potential, while the actual minimum next to it is characterised by the physical mass $m_\phi^2 \simeq m_h^2 \sin^2 \theta$. This means that any point on the naturalness line can be realised in multiple ways, which feature almost identical $\mu$ parameters, but different $m_{\phi_0}$ such that $m_{\phi_0} < \mu$.

As was already mentioned, the best fit value of the $g_{\phi e}$ coupling is in tension with the stellar cooling bounds. 
In the relevant mass range, the strongest constraints are derived from the RGs evolution~\cite{Hardy:2016kme}, $g_{\phi e} \lesssim 10^{-15}$, and are valid for the scalar masses $\lesssim 20\,\keV$.
Such bounds however can be avoided assuming the properties of the $\phi$ field are modified in the dense interior of RG stars. The RG core density significantly exceeds that of the Sun, reaching the nucleon and electron number density $n_{\text{RG}}\sim 10^{15}\,\eV^3$~\cite{Raffelt:1996wa,Hardy:2016kme}, and in principle can affect the local scalar mass, making the cooling bound inapplicable.  
This can be realised for instance if the $\phi$ field potential is characterised by two minima, one being the true minimum in the vacuum, and another becoming the energetically preferred state inside the RG stars, as a result of a correction to the scalar potential $\delta V \simeq g_{\phi N} n_{\text{RG}} \phi$, where $g_{\phi N}$ is a nucleon-scalar coupling. These two minima have to be characterized by significantly different masses. Constructing a potential satisfying all the aforementioned criteria is however a very non-trivial task, which we leave beyond the scope of the current letter.

%%%%%%%%%%%%%%%%%%%%%%%%%%%%%%%%%%%%%%%%
\section{The relaxed relaxion case}
\label{sec:relaxed}
%%%%%%%%%%%%%%%%%%%%%%%%%%%%%%%%%%%%%%%%

\begin{figure}[t]
\vspace{-0.1cm}
\includegraphics[width=0.45\textwidth]{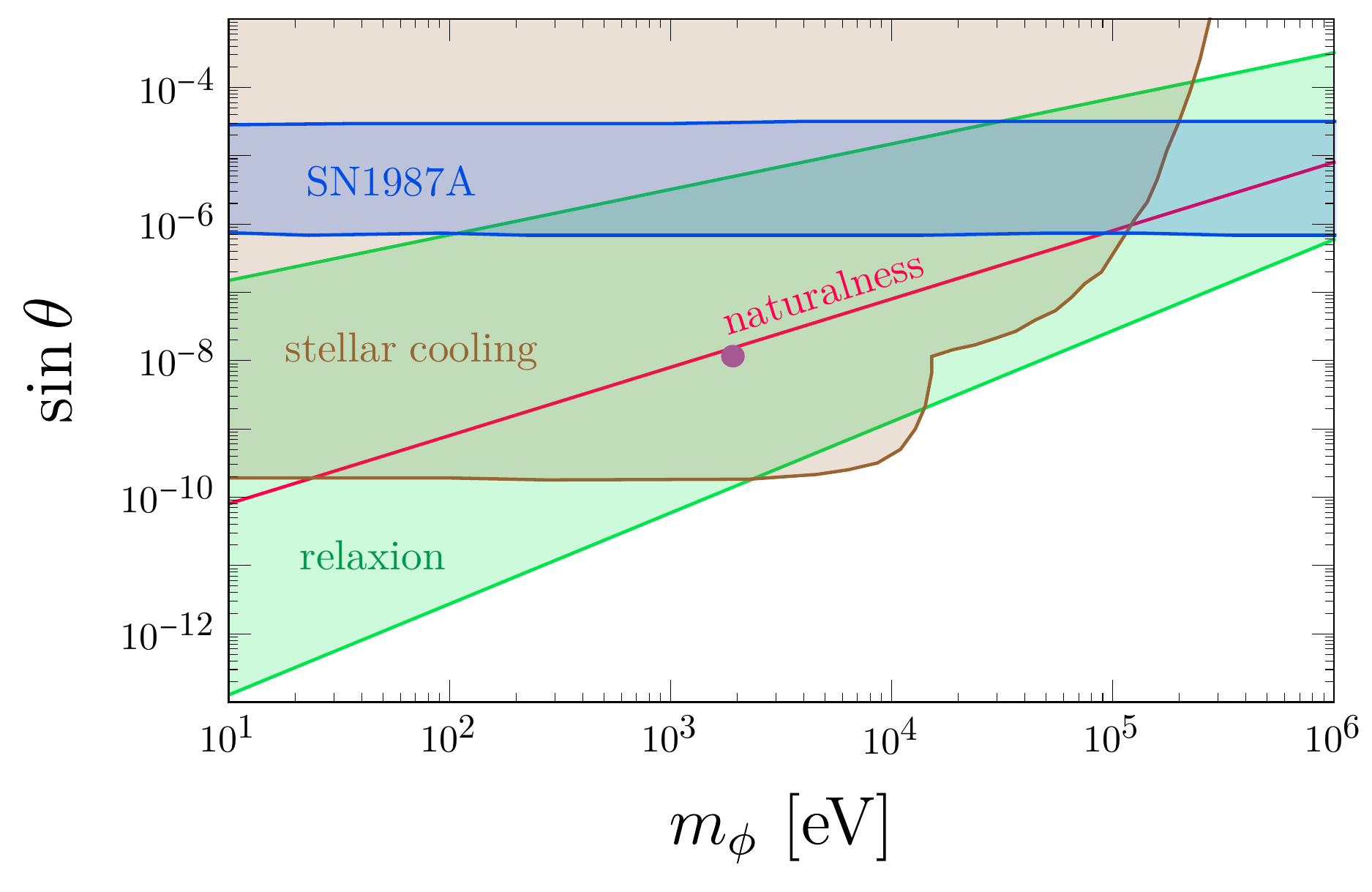}
 \vspace{-0.2cm}
\caption{
The best fit to \XeT excess (purple dot) in terms of the mixing angle, $\sin\theta$, and the scalar mass, $m_\phi$, together with the constraints from the stellar cooling~\cite{Hardy:2016kme,Grifols:1988fv,Raffelt:2012sp}\,(brown), SN1987A~\cite{Turner:1987by,Frieman:1987ui,Burrows:1988ah}\,(blue), as well as preferred relaxion parameter space~\cite{Banerjee:2020kww}\,(green) and the naturalness bound\,(red). 
We note that the SN1987A constraint depends on SNe explosion mechanism~\cite{Bar:2019ifz,RevModPhys.29.547,Fowler:1964zz,Hoyle:1960zz,Kushnir:2014oca}.
}
 \vspace{-0.2cm}
\label{fig:mixmass}
\end{figure}

Relaxion mechanism~\cite{Graham:2015cka} allows to explain the smallness of the Higgs mass by a non-trivial cosmological dynamics of the Higgs-relaxion system. 
The same dynamics also relaxes the relaxion mass from its natural value.
The relaxion mass and electron coupling are predicted~\cite{Banerjee:2020kww} 
\begin{align} \label{eq:relaxmain}
	m_\phi^2 
	\simeq 
	\frac{\Lambda_{\text{br}}^4}{f^2} \frac{\Lambda_{\text{br}}^2}{\Lambda v} \,,\quad
	g_{\phi e} 
= 	\frac{\lambda_e}{\sqrt{2}} \sin \theta \simeq \lambda_e \frac{\Lambda_{\text{br}}^4}{f v^3} \, ,
\end{align} 
where $\Lambda$ is the 
cutoff, $f$ and $\Lambda_{\rm br}$ are the period and amplitude of Higgs-dependent barriers, $\theta$ is the relaxion-Higgs mixing angle, and $v=174\,\GeV$ is the Higgs VEV.  Note that the formulas above are only order of magnitude estimates.
Assuming $f=\Lambda$ and the SM value for the electron Yukawa coupling $\lambda_e = \lambda_{e\text{SM}}$, we find for the best fit values that
\begin{align}
	\Lambda 
= 	f \simeq 60\,\TeV \,,\qquad 
	\Lambda_{\text{br}} 
	\simeq 10\,\GeV \,.\;  
\end{align} 
More generally, for $f>\Lambda$ we have a continuum of possibilities, allowing for $\Lambda < 60\,\TeV <f$ and $\Lambda_{\text{br}}>10\,\GeV$. Furthermore, for $f=\Lambda$, the order of magnitude of the inflationary Hubble parameter is constrained to be within $1\,\eV$ and $0.1\,\GeV$. In Fig.~\ref{fig:mixmass} we show the position of the excess in the allowed parameter space of the relaxion models (green band), together with relevant experimental bounds.

%%%%%%%%%%%%%%%%%%%%%%%%%%%%%%%%%%%%%%%%
%%%%%%%%%%%%%%%%%%%%%%%%%%%%%%%%%%%%%%%% 

For the best fit mass, the relaxion model implies the relaxion-Higgs mixing angle is within the range of $\sin\theta \in [10^{-10},10^{-5}]$~\cite{Banerjee:2020kww}, see Fig.~\ref{fig:mixmass}. 
Thus, by relaxing the assumption of $\lambda_e = \lambda_{e\text{SM}}$, we can identify a preferred range for the electron Yukawa to be $10^{-3} < \lambda_e /\lambda_{e\text{SM}}<10^2$, which is consistent with the current direct upper bound of $\lambda_e \lesssim 600\times \lambda_{e\text{SM}}$~\cite{Khachatryan:2014aep,Altmannshofer:2015qra,Dery:2017axi}.

Finally, we would like to comment on whether the relaxion mechanism can overcome the stellar cooling bounds with a help of the chameleon effect discussed in the previous section\footnote{Density effects on light particles were also considered in other contexts, see e.g.~\cite{Kaplan:2004dq,Hook:2017psm}.}. Potential importance of the density effects inside of neutron stars on the QCD-relaxion was already emphasised in Ref.~\cite{Balkin:2020dsr}, while here we concentrate on the non-QCD version of the relaxion mechanism.
The relaxion potential naturally features a set of consecutive minima, and may travel between them if the density-induced relaxion field displacement is large enough.
In the minimal relaxion scenario, the local nucleon number density $n$ induces a linear piece in the potential $\delta V \simeq g_{\phi N} n \phi$ which shifts the relaxion in the direction of the next deeper minimum. For a sufficiently large shift the relaxion will start rolling towards the next minimum. However, in the absence of efficient friction\footnote{ See~\cite{Hook:2016mqo, Choi:2016kke, Tangarife:2017rgl, Ibe:2019udh, Kadota:2019wyz, Fonseca:2019ypl, Fonseca:2019lmc} for a possible friction source.} and with negligible gradient energy, we expect that the relaxion will not stop until it reaches the global minimum of its potential, featuring a large negative Higgs mass squared of the order of the cutoff scale $\Lambda$. 
If the large density region is larger than the critical bubble, the new phase will expand outside and fill the universe.
Otherwise, localised bubbles~\cite{Hook:2019pbh} within the dense astrophysical objects will be formed.

To induce such a phase transition~(PT), the density-induced relaxion field dispacement, $\delta \phi_n \simeq \delta V'/m_\phi^2 = g_{\phi N} n/m_\phi^2$, has to exceed the distance between the minimum and the closest maximum of the relaxion potential, given by $\Delta \phi \simeq \Lambda_{\text{br}}^2 f/ \Lambda v$~\cite{Banerjee:2020kww}. Using Eq.~(\ref{eq:relaxmain}) we find that the transition requires
\begin{equation} \label{eq:displ1}
	 \frac{\delta \phi_{n}}{\Delta \phi} 
	\simeq \frac{g_{\phi N} n \Lambda^2}{v \Lambda_{\text{br}}^4} \gtrsim 1\,, 
\end{equation} 
and it will be localized within the dense object as long as the object's radius is less than the critical bubble radius which we approximately estimate as $\sim 1/ m_\phi$ (see~\cite{Hook:2019pbh} for a more precise determination of this condition). An existence of such localised phases is an interesting topic which we leave for future studies. 

On the other hand, the scenario with a wrong Higgs VEV bubbles expanding outwards is excluded experimentally. It is important to find out how this fact limits the size of $\sin \theta$ which has a paramount importance for the relaxion experimental tests. 
Expressing Eq.~\ref{eq:displ1} through  $m_\phi$, $\sin \theta$ and $\Lambda$ 
\begin{equation} \label{eq:displ2}
	 \frac{\delta \phi_{n}}{\Delta \phi} 
	\simeq \frac{g_{\phi N} n m_\phi^4 \Lambda^4}{\sin^4 \theta v^{11}} \gtrsim 1\,, 
\end{equation} 
we see that there exists some minimal value of $\sin \theta$, below which the PT always happens, and it is given by
\begin{equation}
\sin \theta_{\text{min}} \simeq \frac{(g_{\phi N} n)^{1/4} m_\phi \Lambda}{v^{11/4}}.
\end{equation}
The absolute lower bound on the mixing is then proportional to the lower bound on the cutoff scale $\Lambda$, for which the mildest estimate would be of order $1$~TeV. 

%%%%%%%%%%%%%%%%%%%%%%%%%%%%%%%%%%%%%%%%
\begin{figure}[t]
\vspace{-0.1cm}
\includegraphics[width=0.45\textwidth]{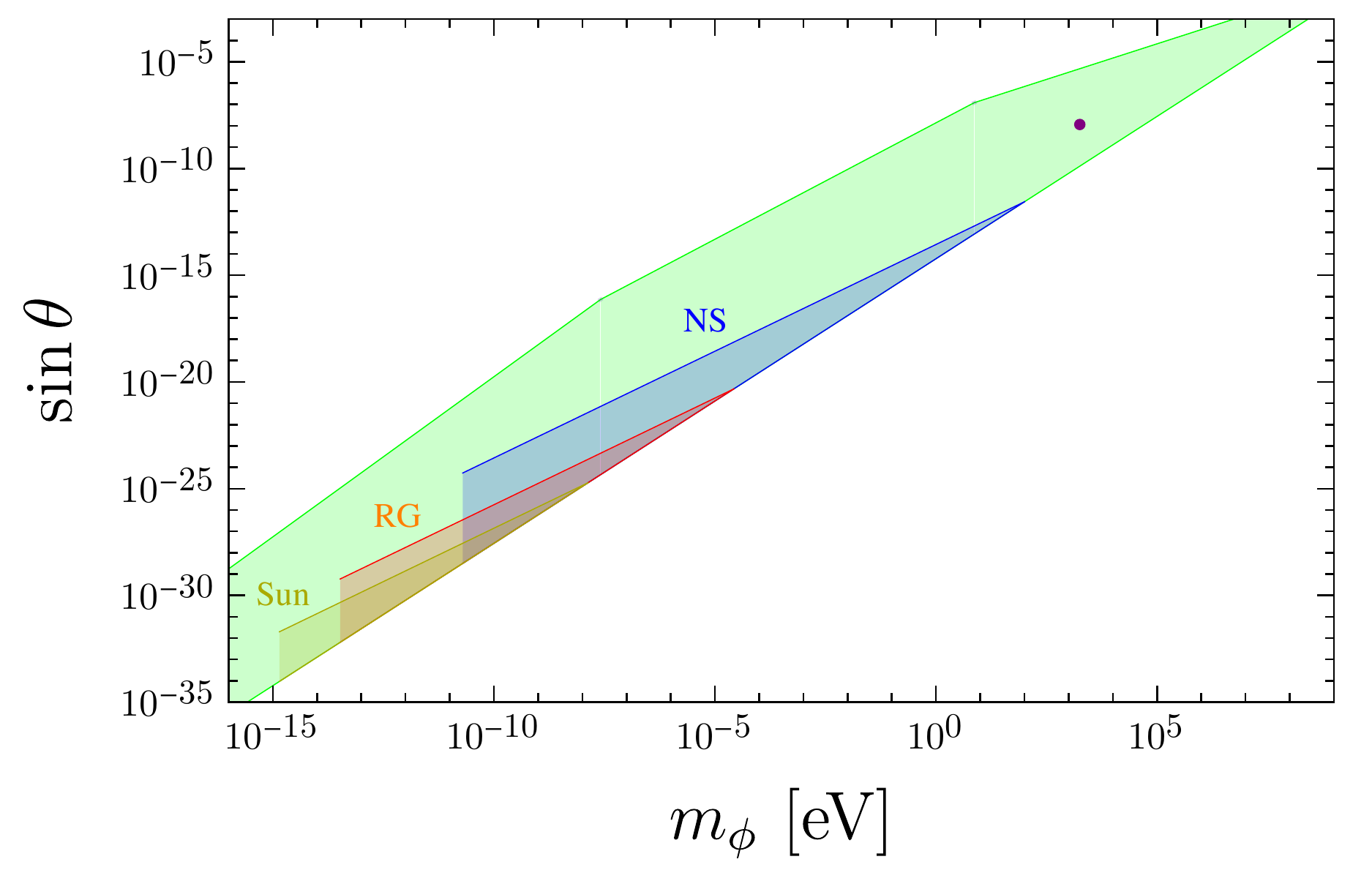}
 \vspace{-0.2cm}
\caption{
Relaxion parameter space in the first minimum~(green) in terms of $\sin\theta$ and $m_\phi$. Purple dot shows the best fit to \XeT excess. 
The blue, red and yellow regions show where the expanding bubbles are produced, resulting from neutron stars, RGs cores and the Sun core, respectively.
See text for more details.}
 \vspace{-0.2cm}
\label{fig:bubbles}
\end{figure} 
%%%%%%%%%%%%%%%%%%%%%%%%%%%%%%%%%%%%%%%% 

In Fig.~\ref{fig:bubbles} we demonstrate our findings, applied to neutron stars, RGs and the Sun. We assume the minimal coupling $g_{\phi N} = g_{h N} \sin \theta$, where $g_{h N} \sim 10^{-3}$ is the Higgs coupling to nucleons.   Colored areas show where the transition always happens and propagates outside the dense objects. Such PT can also happen for larger $\sin \theta$ for some parameter choices. For masses lower than the inverse size of the corresponding astrophysical objects (left to corresponding colored areas) the PT can happen, but it is localised within the dense objects. For this plot we only chose to show the bounds from three distinct types of high-density astrophysical bodies, not aiming at a comprehensive analysis of all possible stars.

As one can see from the plot, the RG-localized PT region is located, as trivially expected, far away from the XENON1T excess point not allowing to reconcile the latter with the stellar cooling.

Notice that the derived bounds can be substantially changed assuming (non-minimal) stronger relaxion coupling to nucleons.
The current experimental bound on proton coupling is $g_{\phi p} \lesssim 10^{-6}$ for $m_\phi\lesssim 0.1$\,keV~\cite{Alighanbari:2020aa}  (for stronger bounds on coupling to neutrons see~\cite{Leeb:1992qf,Nesvizhevsky:2007by,Pokotilovski:2006up,Frugiuele:2016rii}). Such an increased coupling can not help with resolving the stellar cooling tension.

%%%%%%%%%%%%%%%%%%%%%%%%%%%%%%%%%%%%%%%%
\section*{Acknowledgments}
We would like to thank Kfir Blum, Diego Redigolo, and Tomer Volansky for fruitful discussions, and Abhishek Banerjee for technical support.
We are grateful to Oz Davidi for his contributions to this project in its initial phase. 
The work of RB is supported by ISF grant No. 1937/12. RB is the incumbent of the Arye and Ido Dissentshik Career Development Chair.
The work of OM is supported by the Foreign Postdoctoral Fellowship Program of the Israel Academy of Sciences and Humanities.
The work of GP is supported by grants from The U.S.- Israel Binational Science Foundation~(BSF), Israel Science Foundation~(ISF), German Israeli Foundation~(GIF), Yeda-Sela-SABRA-WRC, and the Segre and the Friedrich Wilhelm Bessel Research Awards.
YS is supported by the BSF (NSF-BSF program Grant No. 2018683) and by the Azrieli Foundation.
YS is Taub fellow (supported by the Taub Family Foundation). 
%%%%%%%%%%%%%%%%%%%%%%%%%%%%%%%%%%%%%%%%

%\appendix
 
 \newpage

%%%%%%%%%%%%%%%%%%%%%%%%%%%%%%%%%%%%%%%%
\bibliographystyle{utphys}
\bibliography{XENON_Relaxed_Scalar_bib}
%%%%%%%%%%%%%%%%%%%%%%%%%%%%%%%%%%%%%%%%

\end{document}